\documentclass[aps,pre,superscriptaddress,amsmath,amssymb,reprint,nofootinbib]{revtex4-2}  

\usepackage{graphicx}
\usepackage{mathtools}
\usepackage[hypertexnames=false]{hyperref}
\usepackage{braket}
\usepackage{bm}
\usepackage{xcolor}

\newcommand{\aln}[1]{\begin{align}#1\end{align}}
\newcommand{\nn}{\nonumber\\}

\begin{document}

\title{Revisiting wormhole-induced global symmetry breaking
}

\author{Kiyoharu Kawana}
\email{kkiyoharu@kias.re.kr}
\affiliation{School of Physics, Korea Institute for Advanced Study, Seoul 02455, Korea}


\date{\today}

\begin{abstract}
It is widely believed that global symmetries cannot exist in a consistent theory of quantum gravity.
A prominent mechanism underlying this expectation is provided by gravitational instantons or Euclidean wormholes, whose contributions to the Euclidean path integral generates local symmetry-breaking operators in the effective action with couplings parametrized by the $\alpha$-parameters $\overrightarrow{\alpha}=(\alpha_1^{},\alpha_2^{},\cdots)$. 
The appearance of these $\alpha$-parameters is often taken as evidence that  wormholes explicitly break global symmetries.    
In this paper, we argue that this conclusion is premature. 
%
The fate of the symmetry is determined by the ensemble average over these $\alpha$-parameters. 
We show that, under broad situations, this average can be sharply dominated by the symmetric point $\overrightarrow{\alpha}=0$, while contributions from other symmetry-breaking critical points $\overrightarrow{\alpha}\neq 0$ are typically suppressed by a doubly exponential factor $\exp(-e^{2S_{\rm ins}})$, where $S_{\rm ins}$ is the instanton action. 
In particular, standard $\mathrm{U}(1)^{}$ Peccei--Quinn models fall into this class, implying that 
the conventional axion quality problem does not arise in the wormhole-induced effective theory. 
Our analysis is formulated for general $\mathrm{U}(1)^{}$ $p$-form global symmetries.
\end{abstract}
\maketitle 





\section{Introduction}\label{Sec:intro}

It has long been argued that global symmetries cannot exist in a consistent theory of quantum gravity~\cite{Hawking:1975vcx, Coleman:1988cy, Giddings:1988cx, Kallosh:1995hi, Banks:2010zn, Brennan:2017rbf, Witten:2017hdv, Harlow:2018jwu, Harlow:2018tng, McNamara:2020uza,Yonekura:2020ino}.
A concrete realization of this expectation is provided by Euclidean wormholes or gravitational instantons. 
The Giddings--Strominger wormhole~\cite{Giddings:1987cg} is an Euclidean solution connecting two asymptotically flat spacetimes through an $S^3$ throat supported by the $H_3^{}=dB_2^{}$ flux, where $B_2^{}$ is the dual field of the axion.  
In the dual picture, $\int_{S^3}H_3^{}$ corresponds to the $\mathrm{U}(1)$ Peccei--Quinn (PQ) charge, and the wormhole is transporting the charge from one Universe to another, leading to an explicit breaking of the $\mathrm{U}(1)$ global symmetry. 
%

Summing over such wormholes in the Euclidean path integral generates bilocal terms in the effective action~\cite{Coleman:1988cy, Coleman:1988tj,Klebanov:1988eh,Hebecker:2018ofv,Marolf:2020xie,Witten:2026twr}, and they can be converted into local symmetry-breaking terms by introducing auxiliary $\alpha$-parameters. 
%
These terms correspond to wormhole-induced potentials, whose coupling constants are suppressed by the instanton factor $e^{-S_{\rm ins}^{}}$. 
This factor serves as a qualitative indicator of the strength of symmetry breaking induced by wormholes. 

In this paper, we revisit this wormhole-induced mechanism of global-symmetry breaking by carefully analyzing the ensemble average over the $\alpha$-parameters. 
 We focus on a general matter theory with a $\mathrm{U}(1)$ $p$-form global symmetry and study its effective theory in the presence of the $\alpha$-parameters generated by gravitational instantons. 
Throughout this paper, we denote the $\mathrm{U}(1)$ $p$-form global symmetry simply as $\mathrm{U}(1)^{[p]}$ and the set of $\alpha$-parameters by $\overrightarrow{\alpha }=(\alpha_1^{},\alpha_2^{},\cdots)$.

The key observation is that the mere existence of $\alpha$-parameters does not immediately imply explicit breaking of $\mathrm{U}(1)^{[p]}$; rather, 
the ensemble average can dynamically localize them at zero, depending on the phase structure of the  matter sector.  
More generally, whether global symmetry is truly broken is determined by the location of the critical point in the $\overrightarrow{\alpha}^{}$ space that dominates the ensemble average. 
Of course, this ensemble perspective was already advocated in the Coleman's original  paper~\cite{Coleman:1988tj} for the resolution of the cosmological constant problem, and has been extensively studied in many literatures~\cite{Kawai:2011rj, Kawai:2011qb, Hamada:2014ofa,Hamada:2014xra,Hamada:2015dja,Kawai:2023viy}. 
However, a concrete analysis has been lacking in the context of global-symmetry breaking, particually in Lorentizan systems.

We show that, in the thermodynamic limit $V\to\infty$, the $\overrightarrow{\alpha}$-integral can be sharply dominated by a  critical point of the vacuum energy $\rho(\overrightarrow{\alpha})$~---~either saddle point or quantum phase-transition point~---~while contributions from other critical points are typically suppressed by the doubly exponential factor $\exp(-e^{2S_{\rm ins}^{}})$. 
In particular, depending on the phase structure of the original $\mathrm{U}(1)^{[p]}$-invariant matter sector, the outcome falls into one of the following two cases
\begin{enumerate}
\item When $\mathrm{U}(1)^{[p]}$ is not spontaneously broken, the dominant critical point can be either at $\overrightarrow{\alpha}=0$ or at $\overrightarrow{\alpha}\neq 0$. 
In the former case, $\mathrm{U}(1)^{[p]}$ remains unbroken, wheres it is  explicitly broken in the latter case. 
\item When $\mathrm{U}(1)^{[p]}$ is spontaneously broken, the dominant critical point is at $\overrightarrow{\alpha}=0$. 
Thus, $\mathrm{U}(1)^{[p]}$ is not explicitly broken by wormhole  effects. 
\end{enumerate}
%
%
%
In particular, standard PQ models~($p=0$) fall into the second case, which implies that the axion quality problem does not arise in the wormhole-induced effective theory.

This paper is organized as follows. 
In Sec.~\ref{wormhole theory}, we introduce the general setup and explain how wormholes give rise to  the $\alpha$-parameter ensemble. 
In Sec.~\ref{sec:ensemble average}, we explain the fine-tuning mechanism in general and apply it to the wormhole effective theory.  
In Sec.~\ref{implications}, we discuss several implications of our results. 
In particular, we argue that the axion quality problem does not arise in the present wormhole-induced theory. 
Sec.~\ref{Sec:summary} is devoted to summary.  


\section{Wormhole induced effective theory}\label{wormhole theory}
Here, we discuss the wormhole theory in a general many-body system with a $p$-form global symmetry $\mathrm{U}(1)^{[p]}$ coupled to gravity. 
A typical Euclidean partition function is 
\aln{
\int {\cal D}g\int {\cal D}\phi~e^{-S_G^{}[g]-S_p^{}[\phi,g]}~,
\label{partition function}
}
where $S_G^{}[g]$ is a gravity action and $S_p^{}[\phi,g]$ is a general matter action which possesses $\mathrm{U}(1)^{[p]}$. 
For example, the simplest example is the $p$-form Maxwell theory, i.e., $\phi=A_p^{}=p$-form field, and 
\aln{
S_p^{}[A_p^{},g]=\frac{1}{2g^2}\int_{\Sigma_D^{}}F_{p+1}^{}\wedge \star F_{p+1}^{}~,\quad F_{p+1}^{}=dA_p^{}~. 
}
One can check that this action is invariant under
\aln{
A_p^{}\quad \rightarrow \quad A_p^{}+\theta \Lambda_p^{}~,\quad \theta \in \mathbb{R}~,
}
where $\Lambda_p^{}$ is a nontrivial element of $H^p(C_p^{};\mathbb{Z})$ for a given $p$-dimensional closed subspace $C_p^{}$. 
It is easy to check that the Wilson surface $W[C_p^{}]=\exp\left(i\int_{C_p^{}}A_p^{}\right)$ is charged under this transformation.  

More generally, when a $\mathrm{U}(1)^{[p]}$-invariant matter theory is formulated in terms of local degrees of freedom, one can consider a dual field theory in which the fundamental field $\phi=\phi[C_p^{}]$ is a functional of $p$-dimensional closed spacial manifold $C_p^{}$~\cite{Iqbal:2021rkn,Hidaka:2023gwh,Kawana:2024qmz,Kawana:2025vbi}. 
In this theory, the $p$-form global transformation is expressed by 
\aln{
\label{p-form transformation}
\phi[C_p^{}]\quad \rightarrow \quad e^{2\pi i\theta \int_{C_p^{}}\Lambda_p^{}}\phi[C_p^{}]~,\quad \theta \in \mathbb{R}~,
} 
as the Wilson-surface operator. 
The functional nature of this theory entails numerous gauge redundancies, rendering the analysis of its quantum nature more challenging than in the original local formulation~\cite{Kawana:2025vvf}.  
Nevertheless, a mean-field analysis can be carried out in much the same way as in ordinary scalar field theories, and this is sufficient for our present purpose since our primary interest lies in the phase structure of the system. 
We therefore adapt this effective field-theoretic description in the following discussion.

In this setup, we simply assume the existence of a (brane) instanton solution connecting two asymptotically flat regions and carrying a nonzero $p$-form global charge $m\neq 0$.
This implies that each asymptotic boundary of the solution contains a charged $p$-dimensional brane (or defect) $C_p^{}$, and this is described by the lowest dimension operator
\aln{
\phi[C_p^{}]^m\quad \text{or} \quad \phi[C_p^{}]^{-m}~
}
on each boundary. 
%
%
Then, summing over such wormhole solutions, we obtain the bilocal effective term 
\aln{
c_m^{}e^{-2S_{\rm ins}^{}(m)}\left(\int [dC_p^{}]\phi[C_p^{}]^m\right)\left(\int [dC_p^{'}]\phi[C_p^{'}]^{-m}\right)~,
\label{bi-local action}
}
where $S_{\rm ins}^{}(m)$ is the half instanton action, $c_m^{}$ is a (positive) numerical constant, and $\int [dC_p^{}]$ denotes the path-integral overall embeddings of $p$-brane $C_p^{}$.  
In the following, we simply denote ${\cal O}_m^{}\coloneq \int [dC_p^{}]\phi[C_p^{}]^m$.

The above bilocal term can be cast into a local form by a Gaussian integral as 
 \aln{
e^{c_m^{}e^{-2S_{\rm ins}}{\cal O}_m^{}{\cal O}_{m}^{*}}=\int_{\mathbb{C}}
d^2\alpha 
e^{-|\alpha_m^{}|^2+\alpha_m^{*}\sqrt{c_m^{}}e^{-S_{\rm ins}^{}}{\cal O}_m^{}+{\rm h.c.}}~,
\label{HS transformation}
 }
which then corresponds to a $\mathrm{U}(1)^{[p]}$ breaking potential 
 \aln{
c_m^{}e^{-2S_{\rm ins}}\int [dC_p^{}](\alpha_m^* \phi[C_p^{}]^m+{\rm h.c.})~.
\label{Ap potential}
 }
in the effective action. 
%
In particular, for $p=0$, Eq.~(\ref{Ap potential}) reduces to the wormhole-induced axion potential, whose coefficient is proportional to the instanton suppression factor and has been the subject of considerable recent discussion in connection with the axion quality problem~\cite{Dine:1986bg,Kamionkowski:1992mf,Holman:1992us,Dine:2022mjw,Catinari:2024zon}. 
We will come back to this problem in Section~\ref{implications}. 

More generally, summing over all instantons with different charges and number of legs, we end up with the following effective theory in the Lorentzian (flat) Universe:\footnote{The subscript $M$ here means ``Micro-canonical" since this partition function can be interpreted as  a micro-canonical partition function in statistical mechanics by identifying the $\alpha$-parameters as temperature $T$. 
Instead, one can also interpret it as ``Multiverse". 
}
\aln{
Z_{M}^{}\coloneq &\left(\prod_{m=1}^{\infty}\int_{\mathbb{C}}d^2\alpha_m^{}\right)\omega^{}(\overrightarrow{\alpha})
\nn
\times &\int {\cal D}\phi~\exp\left(iS_p^{}[\phi]+i\left(\sum_{m=1}^\infty e^{-S_{\rm ins}^{}(m)}\alpha_m^{*}{\cal O}_{m}[\phi]+{\rm h.c.}\right)\right)~,
\label{MC ensemble}
}
where we have absorbed $\sqrt{c_m^{}}$ into $\alpha_m^{}$, $S_p^{}[\phi,\eta]=S_p^{}[\phi]$, and $\omega^{}(\overrightarrow{\alpha})$ is a  weight function 
whose functional form is model-dependent. 
Nevertheless, it would be reasonable to assume that it has support in the region $|\overrightarrow {\alpha}|\lesssim 1$ and is exponentially damped for $|\overrightarrow {\alpha}|\rightarrow \infty$, as in Eq.~(\ref{HS transformation}).

\section{average over alpha parameters}\label{sec:ensemble average}
It has been widely argued that the appearance of the $\alpha$-parameters signals the explicit breaking of global symmetries by wormhole effects.  
However, such an argument is incomplete, since it ignores the ensemble average over the $\alpha$-parameters. 
As a result, the precise meaning of global-symmetry breaking remains unclear at this point. 
%

For simplicity, we restrict our attention to a time-independent system and denote its Hamiltonian  by $\hat{H}(\overrightarrow{\alpha})$. 
Equation~(\ref{MC ensemble}) can be then written as 
\aln{
Z_M^{}&=\left(\prod_{m=1}^{\infty}\int d^2\alpha_m^{} \right)\omega(\overrightarrow{\alpha})
\langle f|e^{-i\hat{H}(\overrightarrow{\alpha})T}|i\rangle 
\nn
&=\left(\prod_{m=1}^{\infty}\int d^2\alpha_m^{} \right)\omega(\overrightarrow{\alpha}) e^{-iV\rho(\overrightarrow{\alpha})}
\nn
&\quad \quad \times \sum_{n=0}^{\infty}e^{-iT\Delta E_n^{}(\overrightarrow{\alpha})}\langle f| E_n^{}\rangle \langle E_n^{}|i\rangle
~,
\label{coupling integration}
}
where $|i\rangle$ and $|f\rangle$ are general initial and final states,
$\rho(\overrightarrow{\alpha})$ is the (renormalized) vacuum energy, $|E_n^{}\rangle$ is the $n$-th  energy eigenstate, and $\Delta E_n^{}(\overrightarrow{\alpha})\coloneq E_n^{}(\overrightarrow{\alpha})-E_0^{}(\overrightarrow{\alpha})$ is the corresponding excitation energy. 
%
This shows that the ensemble average is dominantly controlled by the vacuum energy density $\rho(\overrightarrow{\alpha})$ for $V\rightarrow \infty$. 
The same argument applies to correlation functions as discussed in Appendix~\ref{app:expectation value}.

Before turning to the wormhole system, let us consider a simple one-dimensional toy integral in order to grasp the essence.  
Readers familiar with this type of analysis can skip this part and proceed directly to Sec.~\ref{wormhole system}.

\subsection{Fine tuning mechanism in ensemble average}\label{fine tuning}
As a toy case, let us consider the following one-dimensional integral:
\aln{\label{toy model}
\int_{-\infty}^{\infty}d\lambda \omega(\lambda) e^{-iV\rho(\lambda)}~,
} 
where $V$ is supposed to be a large number and $\omega(\lambda )$ is a smooth function of $\lambda$ with finite support. 
As in the wormhole theory, we assume that $\omega(\lambda)$ does not depend on $V$. 
In the limit of $V\rightarrow \infty$, this integral can be sharply dominated by a specific point in the $\lambda$ space as follows. 

\  

\begin{figure}
    \centering
     \includegraphics[scale=0.7]{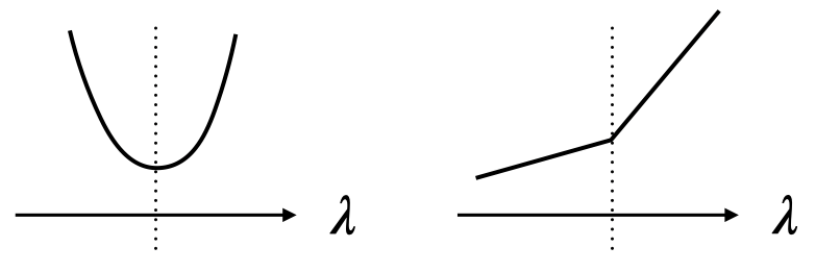}
    \caption{Examples of critical point. 
Left: Saddle point. 
Right: Phase transition point. 
    }
    \label{fig:critical}
\end{figure}

\begin{center}\noindent {\bf Saddle point}
\end{center}
Saddle point is a typical example for such a dominant point, and it is determined by 
\aln{
\frac{\partial \rho(\lambda)}{\partial \lambda}=0~.
}
See the left panel in Fig.~\ref{fig:critical}. 
We denote the solution of this equation generally as $\lambda_c^{}$.  
Around this saddle point, Eq.~(\ref{toy model}) is examined as 
\aln{
&\omega(\lambda_c^{})e^{-iV\rho(\lambda_c^{})}\int_{-\infty}^{\infty}d\lambda~e^{-i\frac{V}{2}\rho''(\lambda_c^{})(\lambda-\lambda_c^{})^2} 
\nn
&\sim~\omega(\lambda_c^{})e^{-iV\rho(\lambda_c^{})}/\sqrt{V\rho''(\lambda_c^{})}~.
}
Namely, we have 
\aln{
e^{-iV\rho(\lambda)}\sim \frac{e^{-iV\rho(\lambda_s^{})}}{\sqrt{V\rho''(\lambda_c^{})}}\delta(\lambda-\lambda_c^{}) 
}
for $V\rightarrow \infty$ as a generalized function.  

\

\begin{center}{\bf Phase transition point}\end{center}
Phase transition point is another example, and we can perform a similar analytic estimation of the integral for  $V\rightarrow \infty$. 
The following argument can be viewed as a Lorentzian counterpart of the standard proof of ensemble equivalence between the micro-canonical and canonical ensembles in statistical mechanics in the presence of a first-order phase transitions at a finite temperature~\cite{10.1119/1.2739571}. 

Assume that $\rho(\lambda)$ is a continuous, smooth and monotonic function except at a point $\lambda=\lambda_c^{}$ at which the first derivative is discontinuous as
\aln{
\Delta_\lambda^{}\coloneq \left(\frac{\partial \rho(\lambda)}{\partial \lambda}\right)^{-1}\bigg|_{\lambda=\lambda_c^{}+0}^{}-\left(\frac{\partial \rho(\lambda)}{\partial \lambda}\right)^{-1}\bigg|_{\lambda=\lambda_c^{}-0}^{}\neq 0~.
} 
See the right panel in Fig.~\ref{fig:critical} for example. 
In this case, by perfuming the partial integration, Eq.~(\ref{toy model}) can be evaluated as 
\aln{
&-\frac{1}{iV}\left[\omega (\lambda)\left(\frac{\partial \rho(\lambda)}{\partial \rho}\right)^{-1}e^{-iV\rho(\lambda)}\right]_{\lambda_c^{}}^{\infty}
\nn
&-\frac{1}{iV}\left[\omega(\lambda)\left(\frac{\partial \rho(\lambda)}{\partial \rho}\right)^{-1}e^{-iV\rho(\lambda)}\right]_{-\infty}^{\lambda_c^{}}+O(V^{-2})
\nn
&=\frac{i}{V}e^{-iV\rho(\lambda_c^{})}\omega(\lambda_c^{})\Delta_\lambda^{}+{O}(V^{-2})~,
}
where we have used $\omega(\pm\infty)=0$. 
This implies 
\aln{
e^{-iV\rho(\lambda)}\sim i\frac{e^{-iV\rho(\lambda_s^{})}}{V}\Delta_\lambda^{}\delta(\lambda-\lambda_s^{})
}
for $V\rightarrow \infty$ as a generalized function. 
As a particular case, one can consider $\rho(\lambda)$ which is smooth, monotonic, and defined only  on the half line $[\lambda_0^{},\infty)$ or $(-\infty,\lambda_0^{}]$. 
In this case, the boundary point $\lambda_0^{}$ dominates the integral. 
Of course, these results are immediate consequences of the Riemann--Lebesgue theorem. 
This argument can be straightforwardly generalized to higher-order phase transitions, where higher derivative of $\rho(\lambda)$ has a discontinuous point. 

How is this fine-tuning mechanism related to global symmetry ? 
In quantum many-body systems, a quantum phase-transition point $\lambda_c^{}$ often correspond to a point at which the systems undergoes spontaneous symmetry breaking. 
 It is therefore natural to expect that a symmetry-enhanced point may serve as the dominant critical point in the ensemble average.   
Moreover, when this fine-tuning mechanism operates, non-local effects are largely suppressed by negative powers of the spacetime volume $V$.  
In other words, locality is effectively restored in the thermodynamic limit. 

\subsection{The wormhole system}\label{wormhole system}

With these preparations in place, let us now turn to the wormhole effective theory~(\ref{MC ensemble}). 
Since each wormhole-induced term is suppressed by the factor $e^{-S_{\rm ins}^{}(m)}\simeq (e^{-S_{\rm ins}^{}(1)})^m$, we neglect the higher-order contributions ($m\geq 2$) for the moment and focus on the leading $m=1$ term. 
See Appendix~\ref{app:higher order} and the last two paragraphs of this section about the effects of these higher-order terms.  

\begin{figure}
    \centering
     \includegraphics[scale=0.7]{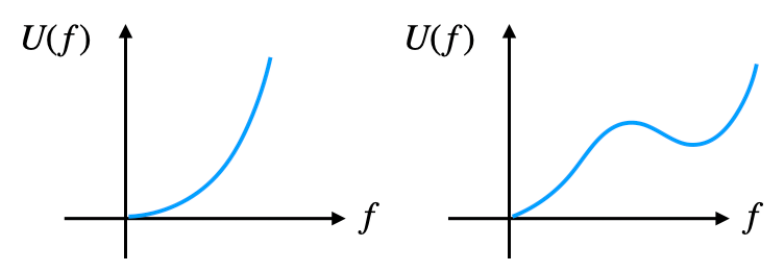}
 \includegraphics[scale=0.75]{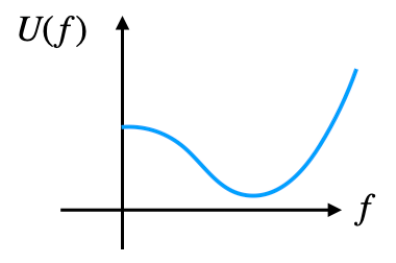}
    \caption{Typical shapes of the bare potential $U(f)$.  
    }
    \label{fig:potential}
\end{figure}

Within this approximation, the canonical partition function reduces to
\aln{
Z_{C}^{}(\lambda)\sim \int {\cal D}\phi~\exp\left(iS_p^{}[\phi]-i\lambda \int [dC_p^{}](\phi+\phi^*)\right)~,
\label{reduced partition function}
}
where $\lambda=\sqrt{c_1^{}}e^{-S_{\rm ins}^{}(1)}|\alpha_1^{}|$ and we have absorbed the phase of $\alpha_1^{}$ into $\phi[C_p^{}]$.  
%
Introducing the polar expression $\phi=f\times e^{i\theta}~,~f\geq 0~,\theta \in [0,2\pi)$, the effective potential of $\phi$ is written as 
\aln{
U^{}(f)+\lambda f\cos (\theta)~,
}
where $U^{}(f)$ is the (bare) potential in $S_p^{}[\phi]$, which is assumed to be bounded below.  
Since $\lambda f\geq 0$, $\theta$ has a global minimum at $\theta=\pi$, and we obtain the effective potential of $f$ as 
\aln{
U_{\rm eff}^{}(f)\coloneq U(f)-\lambda f~.
\label{effective potential of f}
}
This potential can have an extremum determined by  
\aln{
\frac{dU}{df}-\lambda=0~,
\label{vacuum}
} 
whose location depends on the shape of the bare potential $U(f)$. 
Here, there are typically three possibilities:  
\begin{enumerate}
\item $U(f)$ has a global minimum at $f=0$ and no other local extrema. 
This is shown in the upper left panel in Fig.~\ref{fig:potential}. 
\item $U(f)$ has a global minimum at $f=0$, but also has other local minima, as shown in the upper right panel in Fig.~\ref{fig:potential}. 
\item $U(f)$ has a global minimum at $f=v\neq 0$, i.e., $\mathrm{U}(1)^{[p]}$ is spontaneously broken in the original system. 
This is shown in the lower panel in Fig.~\ref{fig:potential}. 
\end{enumerate}
In the first case, there exists an unique solution $f=v(\lambda)$ of Eq.~(\ref{vacuum}) for $^\forall \lambda\geq 0$. 
Correspondingly, the vacuum energy $\rho(\lambda)=U_{\rm eff}^{}(v(\lambda))$ is a monotonically decreasing function of $\lambda$. 
%

In a similar manner, one can see that the vacuum energy $\rho(\lambda)$ is a monotonically decreasing function of $\lambda$ in the third case because the linear term $-\lambda f$ simply  lowers the existing minimum of the potential.   
Aa a result, by the fine-tuning mechanism discussed in Sec.~\ref{fine tuning}, we find 
\aln{
\log Z_M^{}=\log Z_C^{}(\lambda=0)+{\cal O}(\log V)
\label{equivalence}
}
in both first and third cases. 
Equation~(\ref{equivalence}) implies that, in the thermodynamic limit, the wormhole effective theory~(\ref{reduced partition function}) is equivalent to the ordinary canonical theory with no explicit breaking of $\mathrm{U}(1)^{[p]}$. 

The second case is more nontrivial compared to the above two cases.  
In this case, there exists a critical point $\lambda=\lambda_c^{}\neq 0$ at which two minima become  degenerate as 
\aln{
U_{\rm eff}^{}(0)=U_{\rm eff}^{}(v(\lambda_c^{})).
}
Correspondingly, the vacuum energy behaves as  
\aln{
\rho(\lambda)=\begin{cases}
U_{\rm eff}^{}(0) & \text{for } 0\leq \lambda \leq \lambda_c^{}
\\
U_{\rm eff}^{}(v(\lambda)) & \text{for } \lambda_c^{}\leq \lambda 
\end{cases}
}  
as a function of $\lambda$. 
This means that $\lambda=\lambda_c^{}$ is a first-order phase transition point, and we obtain the ensemble equivalence 
\aln{
\log Z_M^{}=\log Z_C^{}(\lambda =\lambda_c^{})+{\cal O}({\log V})
}  
by the fine-tuning mechanism in Sec.~\ref{fine tuning}. 
Here, $\mathrm{U}(1)^{[p]}$ remains explicitly broken even after averaging over the coupling constant. 

In summary, we have shown that the point 
\aln{
\overrightarrow{\alpha}_c^{(0)}=(\alpha_c^{},0,0,\cdots )~,\quad \alpha_c^{}=0~\text{or}~\lambda_c^{}e^{S_{\rm ins}^{}(1)}
}
is at least one critical point that contributes to the ensemble average~(\ref{MC ensemble}). 
In general, in the presence of higher-order terms, additional critical points $\overrightarrow{\alpha}_c^{(k)}~(k=1,2,\cdots)$ can appear with $\alpha_{m,c}\neq 0$ for some $ m\geq 2$.  
Then, the expectation value of a general observable ${\cal O}$ is given by 
\aln{
\label{expectation value}
\langle {\cal O}\rangle_M^{}=
\langle {\cal O}\rangle_C^{(0)}+\sum_{k=1,2,\cdots }\frac{\omega(\overrightarrow{\alpha}_c^{(k)})}{\omega(\overrightarrow{\alpha}_c^{(0)})}\langle {\cal O}\rangle_C^{(k)}~,
}
where $\langle {\cal O}\rangle_C^{(l)}$ denotes the canonical expectation value evaluated at the coupling constants  $\overrightarrow{\alpha}=\overrightarrow{\alpha}_c^{(l)}$. 
These additional contributions are, however, typically suppressed 
because a nonzero critical coupling $\alpha_{m,c}$ is often proportional to a positive power of $e^{S_{\rm ins}^{}(1)}$, which consequently gives rise to a doubly exponential factor 
\aln{
\omega(\overrightarrow{\alpha}_c^{(k)})\propto e^{-|\alpha_{m,c}^{}|^2}\sim \exp\left(-e^{2S_{\rm ins}^{}(1)}\right)
}
See Appendix~\ref{app:higher order} for a concrete example of such an additional critical point.    

Of course, one can in principle realize additional critical points satisfying $\omega(\overrightarrow{\alpha}_c^{(m)})\sim  \omega(\overrightarrow{\alpha}_c^{(0)})$ by appropriately choosing the original potential $U(f)$. 
In such a case, the expectation value~(\ref{expectation value}) receives  multiple comparable  contributions from these critical points. 
This should be interpreted as an average over a multiverse of theories.

\section{Implications}\label{implications}
In this section, we discuss several implications of the results presented in the previous section.  
\subsection{Axion quality problem}
The first implication concerns the axion quality problem~\cite{Dine:1986bg,Kamionkowski:1992mf,Holman:1992us,Dine:2022mjw,Catinari:2024zon}. 
Axion corresponds to the $p=0$ case, and Eq.~(\ref{MC ensemble}) leads to the axion potential    
\aln{
\sum_{m=1}^{\infty}|c_m^{}|(f_a^{})^m\cos\left(m\frac{a}{f_a^{}}+\delta_m^{}\right)~
\label{axion potential}
}
where $f_a^{}$ is the axion decay constant, and $|c_m^{}|\propto e^{-S_{\rm ins}^{}(m)}|\alpha_m^{}|$ is the effective positive coupling constant. 
The presence of these corrections shifts the location of the global minimum of the axion potential and induces a nonzero effective CP phase, $\overline{\theta}\neq 0$~\cite{Abel:2020pzs}.
Consequently, the experimental bound $|\overline{\theta}|\lesssim 10^{-10}$ places a strong constraint on $|c_m^{}|$.

In the wormhole-induced theory~(\ref{MC ensemble}), the coupling constant $|c_m^{}|$ is proportional to the instanton suppression factor $e^{-S_{\rm ins}^{}(m)}$.
Therefore, the experimental bound can be avoided if $S_{\rm ins}^{}(1)\gtrsim 200$ for $f_a^{}\lesssim 10^{16}~$GeV~\cite{Kallosh:1995hi}. 
This condition is, however, nontrivial to realize in standard PQ models since the radial component of the PQ scalar provides a large negative contribution to $S_{\rm ins}^{}(1)$. 
See Refs.~\cite{Hamaguchi:2021mmt,Cheong:2023hrj,Cheong:2024kum,Catinari:2024zon} and references therein for further  details and possible resolutions by non-minimal extensions.   

On the other hand, the results in the previous section provide an alternative and natural solution to the axion quality problem:  
A typical PQ model contains a PQ complex scalar $\phi=f\times e^{i\theta}$ with a potential $U(f)$ exhibiting the spontaneous breaking of the PQ $\mathrm{U}(1)$ symmetry. 
This setup corresponds the third case in Sec.~\ref{wormhole system}. 
Thus, the symmetric point $\overrightarrow{\alpha}=\overrightarrow{\alpha}_c^{(0)}=(0,0,\cdots)$ is  the critical point that gives a dominant contribution in the ensemble average, and   contributions from other critical points are negligibly small because they are suppressed by the doubly exponential factor $\exp\left(-e^{2S_{\rm ins}^{}(1)}\right)$. 
%
%
As a result, the axion quality problem does not arise in standard PQ models in the present wormhole induced effective theory. 
Note that this mechanism is reminiscent of the Coleman's original resolution to the cosmological constant problem~\cite{Coleman:1988tj}. 
%

\subsection{Baby universe hypothesis}
The baby universe hypothesis~\cite{McNamara:2020uza,Marolf:2020xie} asserts that the Hilbert space ${\cal H}_{\rm BU}^{}$ of baby universes must be one-dimensional. 
This conjecture is supported by several theoretical considerations. 
For example, one simple motivation comes from the potential loss of information due to baby universes:  
Suppose that the combined system of the effective theory and the baby universe is initially a pure state such as $|i_{\rm BU}^{}\rangle \otimes |i_{\rm EFT}^{}\rangle $. 
Under time evolution, the system evolves to an entangled state 
\aln{
\sum_{i=1}^{\mathrm{dim}{\cal H}_{\rm BU}^{}} c_i^{}|\alpha_i^{}\rangle _{\rm BU}^{}\otimes |\psi_{\rm EFT}^{};\alpha_i^{}\rangle~,
\label{entangled state}
}
through general interactions, where the eigenvalue $\alpha_i^{}$ corresponds to the $\alpha$-parameter. 
After tracing out the baby universe sector, this leads to the mixed state 
\aln{
\sum_{i=1}^{\mathrm{dim}{\cal H}_{\rm BU}^{}}|c_i^{}|^2|\psi_{\rm EFT}^{};\alpha_i^{}\rangle \langle \alpha_i^{};\psi_{\rm EFT}^{}|~.
}
Thus, if $\mathrm{dim}{\cal H}_{\rm BU}^{}\geq 1$, information is  effectively  lost from the point of view of the effective theory.  
On the other hand, when $\mathrm{dim}{\cal H}_{\rm BU}^{}=1$, Eq.~(\ref{entangled state}) is essentially a tensor-product state, and no such information loss occurs. 
See Ref.~\cite{McNamara:2020uza} and references therein for more details and also other motivations, including arguments based on Swampland conjectures.

Our results in the previous section also supports this baby universe hypothesis but from a different perspective. 
We argue that even when $\mathrm{dim}{\cal H}_{\rm BU}^{}>1$, Eq.~(\ref{entangled state}) can be effectively reduces to
\aln{
\sim |\alpha_{c}^{}\rangle _{\rm BU}^{}\otimes |\psi_{\rm EFT}^{};\alpha_c^{}\rangle~
}
in the thermodynamic limit, where $\alpha_c^{}$ is the critical eigenvalue determined by the vacuum energy of the effective theory.  
In this sense, the ensemble average dynamically selects a single $\alpha$-state, and the effective theory does not suffer from information loss associated with superpositions of different baby universe sectors. 
%

\subsection{Multi-critical point principle}
\begin{figure}
    \centering
    \includegraphics[scale=0.8]{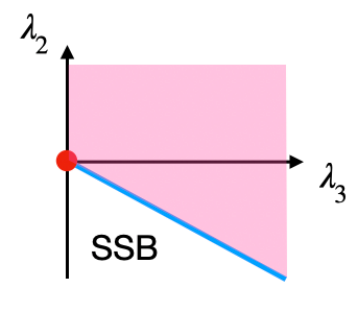}
    \caption{Phase diagram obtained by the potential~(\ref{scalar potential}). 
    }
    \label{fig:phase_diagram}
\end{figure}

The fine-tuning mechanism discussed in Section~\ref{sec:ensemble average} is not entirely new; a closely related idea has long been known as the multi-critical point principle (MPP) in the literatures~\cite{Froggatt:1995rt,Nielsen:2012pu,Hamada:2015dja,Hamada:2017yji,Kawai:2021lam,Kawai:2023viy}. 
In the MPP, one considers a low-energy effective theory in a generalized ensemble such as 
\aln{
\label{general ensamble}
\Omega^{}&\coloneq \int {\cal D}\phi~W({\cal O}_1^{}[\phi],{\cal O}_2^{}[\phi],\cdots )
\\
&=\left(\prod_{k}\int_{-\infty}^{\infty}\frac{d\lambda_k^{}}{2\pi}\right)w(\overrightarrow{\lambda})\int {\cal D}\phi~e^{i\sum_{k}\lambda_k^{}{\cal O}_k^{}[\phi]}~,
} 
where ${\cal O}_k^{}[\phi]$ denotes the spacetime integral of a local operator, $W(x_1^{},x_2^{},\cdots)$ is a general weight function, and $w(\overrightarrow{\lambda})$ is the Fourier transform of it.\footnote{
In particular, the micro-canonical ensemble corresponds to $W({\cal O}_1^{}[\phi],{\cal O}_2^{}[\phi],\cdots )=\delta({\cal O}_1^{}[\phi]-A_1^{})\delta({\cal O}_2^{}[\phi]-A_2^{})\times \cdots$, where $A_k^{}$ is an extensive parameter corresponding to ${\cal O}_k^{}[\phi]$. 
}
This partition function is essentially of the same form as the wormhole induced ensemble~(\ref{MC ensemble}), and the coupling constants $\overrightarrow{\lambda}$ are fixed at the critical point that dominates the ensemble average. 

If all possible low-energy operators can be included in Eq.~(\ref{general ensamble}), it is natural to expect that a symmetry-enhanced point appears as a critical point. 
For example, consider a real scalar field $\phi$ with the renormalizable potential
\aln{
\label{scalar potential}
U(\phi)=\frac{\lambda_2^{}}{2}\phi^2+\frac{\lambda_3^{}}{3!}\phi^3+\frac{\lambda_4^{}}{4!}\phi^4~.
}
For simplicity, we fix $\lambda_4^{}>0$ and regard only $\lambda_2^{}$ and $\lambda_3^{}$ as   ensemble parameters.  
In this case, the phase diagram can be obtained straightforwardly, as shown in Fig.~\ref{fig:phase_diagram}. 
Here, each point on the blue line corresponds to a first-order phase transition point, and the symmetric point $\lambda_2^{}=\lambda_3^{}=0$ (red point) is a second-order phase transition point. 
By applying the fine-tuning mechanism in Sec.~\ref{fine tuning}, one finds that the symmetric point dominates the ensemble average, leading to the emergence of $\mathbb{Z}_4^{}$ in this case. 
In this way, symmetry-enhanced points provide typical critical points in the general ensemble~(\ref{general ensamble}).
Besides, this also implies that the quadratic divergence problem can be absent in this generalized formulation~\cite{Kawai:2023viy}.\footnote{
If one includes quantum loop corrections, $\lambda_2^{}$ should be interpreted as the renormalized mass, and its vanishing point corresponds to the second-order phase transition point, which is naturally preferred in the generalized ensemble.   
}

%
%
%

\section{Summary}\label{Sec:summary}

In this paper, we have revisited global-symmetry breaking induced by wormholes. 
%
%
%
We argued that the integral over the $\alpha $-parameters is governed by a critical point of the vacuum energy $\rho(\overrightarrow{\alpha})$, which may be an ordinary saddle point or a quantum phase-transition point. 
In particular, the symmetric point $\overrightarrow{\alpha}=0$ can often emerge as  the dominant  critical point since the matter sector often exhibits a phase transition there.  
%
%
When multiple critical points exist, contributions from other critical points are typically suppressed by the doubly exponential factor $\exp(-e^{2S_{\rm ins}})$ rather than by the usual instanton factor.
These observations refine the conventional interpretation of wormhole-induced symmetry breaking: the fate of global symmetry is determined by the vacuum structure of the matter sector and the critical point that dominates in the $\alpha$-parameter ensemble. 

As an important application, we showed that standard PQ models fall into the class of theories in which the $\alpha$-integral is dominated by the symmetric point. 
This implies that the conventional axion quality problem does not arise in the present wormhole-induced effective theory.  

%
%

\section*{Acknowledgements} 
We would like to thank Yuta Hamada for fruitful comments. 
This work is supported by KIAS Individual Grants, Grant No. 090901.

\appendix 
\section{Correlation functions}\label{app:expectation value}
We can also check the equivalence of correlation functions. 
As usual, we introduce a source term as 
\aln{
Z_M^{}[J]
=&\left(\prod_{m}\int d^2\alpha_m^{}\right)\omega(\overrightarrow{\alpha}) 
\nn
&\times \int {\cal D}\phi~\exp\left(iS[\phi]+i\int d^Dx J(x)\phi(x)\right)
\nn
&=\left(\prod_{m}\int d^2\alpha_m^{}\right)\omega(\overrightarrow{\alpha})Z_C^{}(\overrightarrow{\alpha})G[J;\overrightarrow{\alpha}]~,
}
where 
\aln{
G[J;\overrightarrow{\alpha}]\coloneq \frac{1}{Z_C^{}(\overrightarrow{\alpha})}\int {\cal D}\phi~\exp\left(iS[\phi]+i\int d^Dx J(x)\phi(x)\right),
}
is the generating functional for (non-connected) correlation functions in the canonical ensemble.  
As long as $J(x)$ is a finite supported function, the factor $G[J;\overrightarrow{\alpha}]$ does not have exponentially large volume dependence, and the integral over the $\alpha$-parameters is still dominated by the critical point of $Z_C^{}(\overrightarrow{\alpha})$.
As a result, by taking the functional derivatives, we obtain 
\aln{
&\langle T\{\phi(x_1^{})\phi(x_2^{})\cdots \phi(x_n^{})\}\rangle_M^{}
\nn
=&\langle T\{\phi(x_1^{})\phi(x_2^{})\cdots \phi(x_n^{})\}\rangle_C^{}\bigg|_{\overrightarrow{\alpha}=\overrightarrow{\alpha}_c^{}}^{}~
}
in the thermodynamic limit, where $T$ means the time-ordered product and $\overrightarrow{\alpha}_c^{}$ is the dominant critical point.

\section{Example of nontrivial critical point}\label{app:higher order}

We generally denote the effective potential as
\aln{
U(f)+\Delta U(f;\overrightarrow{\alpha})~,
}
where $\Delta U(f;\overrightarrow{\alpha})$ is the sum of  all wormhole induced terms. 
As usual, we choose the global minimum of the potential zero. 
Besides, let us denote the VEV of nontrivial local minimum generally by $v(\overrightarrow{\alpha})$. 
Then, critical point is determined by the condition 
\aln{
U(v(\overrightarrow{\alpha}_c^{}))+\Delta U(v(\overrightarrow{\alpha}_c^{});\overrightarrow{\alpha}_c^{})=0~,
\label{critical point condition}
} 
which can be interpreted as the balance condition between the original bare potential $U(f)$ and the wormhole induced term $\Delta U(f;\overrightarrow{\alpha})$.  
This can be realized only when the higher-order couplings $\{\alpha_m^{}\}_{m\geq 2}^{}$ compensate the suppression factors $e^{-S_{\rm ins}^{}(m)}$ in $\Delta U(v(\overrightarrow{\alpha}_c^{});\overrightarrow{\alpha}_c^{})$. 

As a concrete example, let us consider the following toy model:
\aln{
&U(f)=\lambda (f^2-1)^2+\kappa f^6~,\quad \lambda~,~\kappa >0~,
\\
&\Delta U(f;\alpha_5^{})=-e^{-S_{\rm ins}^{}(5)}\alpha_5^{}f^5~.
}
As long as $\lambda \gg \kappa$, $U(f)$ has a global minimum at $f\simeq 1$, and another local minimum can appear when $\alpha_5^{}$ is sufficiently large.  
Qualitatively, the corresponding VEV  can be estimated as  
\aln{
e^{-S_{\rm ins}^{}(5)}\alpha_5^{}v^5\sim \kappa v^6\quad \therefore~v(\alpha_5^{})\sim e^{-S_{\rm ins}^{}(5)}\alpha_5^{}/\kappa~,
} 
which leads to  
\aln{
&U(f)+\Delta U(f;\alpha_5^{})|_{f=v(\alpha_5^{})}^{}
\nn
\sim &~\lambda (e^{-S_{\rm ins}^{}(5)}\alpha_5^{}/\kappa)^4-(e^{-S_{\rm ins}^{}(5)}\alpha_5^{})^6/\kappa^5~.
}
Now, the condition for the critical point (\ref{critical point condition}) is solved as 
\aln{
\alpha_5^{}\sim \lambda \kappa e^{+S_{\rm ins}^{}(5)}~,
}
which confirms the general argument. 
In this way, contributions from other nontrivial critical points are typically suppressed by  
\aln{
\omega (\overrightarrow{\alpha}_c^{})\sim \exp\left(-|\overrightarrow{\alpha}_c^{}|^2\right)\sim \exp\left(-e^{2mS_{\rm ins}^{}(1)}\right)
} 
where $m$ is a positive integer. 
%

\bibliography{Bibliography}

\end{document}